\documentclass[twoside,twocolumn]{article}

\usepackage{blindtext} % Package to generate dummy text throughout this template 

\usepackage{graphicx}
\usepackage{subfigure}

\usepackage[sc]{mathpazo} % Use the Palatino font
\usepackage[T1]{fontenc} % Use 8-bit encoding that has 256 glyphs
\usepackage{microtype} % Slightly tweak font spacing for aesthetics

\usepackage[english]{babel} % Language hyphenation and typographical rules

\usepackage[hmarginratio=1:1,top=32mm,columnsep=20pt]{geometry} % Document margins
\usepackage[hang, small,labelfont=bf,up,textfont=it,up]{caption} % Custom captions under/above floats in tables or figures
\usepackage{booktabs} % Horizontal rules in tables

\usepackage{lettrine} % The lettrine is the first enlarged letter at the beginning of the text

\usepackage{enumitem} % Customized lists
\setlist[itemize]{noitemsep} % Make itemize lists more compact

\usepackage{abstract} % Allows abstract customization
\usepackage{titlesec} % Allows customization of titles
\renewcommand\thesection{\Roman{section}} % Roman numerals for the sections
\renewcommand\thesubsection{\roman{subsection}} % roman numerals for subsections
\titleformat{\section}[block]{\large\scshape\centering}{\thesection.}{1em}{} % Change the look of the section titles
\titleformat{\subsection}[block]{\large}{\thesubsection.}{1em}{} % Change the look of the section titles

\usepackage{fancyhdr} % Headers and footers
\usepackage{titling} % Customizing the title section

\usepackage{hyperref} % For hyperlinks in the PDF

\newcommand{\rta}{\rightarrow}

\newcommand{\beq}{\begin{equation}}
\newcommand{\eeq}{\end{equation}}

\newcommand{\ball}{\begin{align}}
\newcommand{\eall}{\end{align}}

\newcommand{\beqar}{\begin{eqnarray}}
\newcommand{\eeqar}{\end{eqnarray}}

\newcommand{\ben}{\begin{enumerate}}
\newcommand{\een}{\end{enumerate}}

\pretitle{\begin{center}\huge\bfseries} % Article title formatting
\posttitle{\end{center}} % Article title closing formatting
\title{A "road-map" of Nickelate superconductivity} % Article title
\author{%
\textsc{Navinder Singh}\thanks{Cell Phone: +919662680605} \\[1ex] % Your name
\normalsize Physical Research Laboratory, Ahmedabad, India. \\ % Your institution
\normalsize \href{mailto:navinder.phy@gmail.com}{navinder.phy@gmail.com} % Your email address
}
\date{\today} % Leave empty to omit a date
\renewcommand{\maketitlehookd}{%
\begin{abstract}
We comment on the electronic structure of Nickelate system $Nd Ni O_2$ which shows superconductivity on doping. The doped system $Nd_{0.8}Sr_{0.2}NiO_2$ shows (most probably unconventional) superconductivity with transition temperatures ranging from $9~K$ to $15~K$\cite{dang}. The resistivity of the parent compound $Nd NiO_2$ shows a minimum at around 60K, and an up-turn at lower temperatures. In the current understanding it is thought to be originating from a Kondo like coupling between mobile d-states of the $Nd~5d$ bands and un-paired electrons in $Ni 3d_{x^2-y^2}-O2p$ hybrid orbitals of $Ni^{1+}$ and $O^{2-}$ions.  In this note we offer alternative possibilities of the observed resistance minimum in $LaNiO_2$. We attempt to construct a "road-map" of various possibilities to comprehend the observed behaviour and suggest further experiments to prove or disprove a given explanation. We also compare and contrast Nickelates  with Cuprates. Comparison points towards unconventional superconductivity in Nickelates.  
\end{abstract}
}

\begin{document}

% Print the title
\maketitle

%----------------------------------------------------------------------------------------
%	ARTICLE CONTENTS
%----------------------------------------------------------------------------------------

\section{Introduction}
Recent discovery of $9~K$ to $15~K$ superconductivity in the compound $Nd_{1-x}Sr_xNiO_2$\cite{dang} has created much excitement and electronic structure calculations has been reported\cite{yus,hep,hir,hiro,bot,xia}. The parent compound $NdNiO_2$ is a conductor which shows peculiar resistivity (In contrast, parent compounds of Cuprates are magnetic insulators). Formal valence counting in $NdNiO_2$ will suggest $Ni$ to be in $d^9$ state\footnote{$Ni: [Ar] 3d^8 4s^2$ and $Nd: [Xe]4f^4 5d^0 6s^2$, in the compound we will have $Nd^{3+}([Xe]4f^3)~~Ni^{1+}([Ar]3d^9)~~[O^{2-}([He]2s^2 2p^6)]_2$.} similar to $d^9$ configuration of Cu in Cuprates.

This similarity was recognized roughly two decades ago\cite{ani} where authors investigated possible nickelate analogs to the Cuprates. They recognized that $Ni^{1+}$ is a difficult valence and suggested (based on LDA+U calculations) that $LaNiO_2$ will be a magnetic insulator (with Ni having rare $d^9$ states and un-paired electrons in $Ni 3d_{x^2-y^2}$ state having dominating weight at the Fermi level). On doping it should lead to (S=0) $Ni^{2+}$ states imbedded in the matrix of (S=1/2) $Ni^{1+}$ states. Similarity with cuprates was clear, however, authors recongnized the very different local environment. They also studied other valence states of $Ni$  ($Ni^{2+}$ and $Ni^{3+}$).

In $NiO$ and in $La_2NiO_4$, $Ni$ is in the usual valence state of $2+$ that is $d^8$ configuration. These compounds form insulators. $Ni$ can be in the valence state $Ni^{3+}$ as in $LaNiO_3$ ($La^{3+}Ni^{3+}[O^{2-}]_3$). Formally, $Ni$ should have $d^7$ configuration ($Ni:[Ar]3d^84s^2 ~and~Ni^{3+}[Ar]3d^7$), however, it turns out that due to "self-doping" holes are created in oxygen $p$ orbitals ($Ni^{3+}O^{2-}\rta Ni^{2+}O^{1-}$) which constitute mobile states and localized (S=1) states are formed on $Ni^{2+}$ (one un-paired electron in $Ni3d_{x^2-y^2}$ orbital and another unpaired electron in $Ni3d_{3z^2-r^2}$ orbital). It was pointed out that there exists strong coupling between mobile $p$ holes with localized $Ni$ spins via $2p-3d$ hybridization and the system ($LaNiO_3$) should lead to heavy fermion Kondo type behaviour which was later found to be so\cite{dang, ani}.

Authors in the recent discovery\cite{dang} reduce $LaNiO_3$ to $LaNiO_2$ via difficult low temperature topochemical reaction ("Jenga chemistry")\cite{org}. However, it turns out that resistivity of $LaNiO_2$ shows insulating behaviour below 150 K. Then, the authors\cite{dang} explored $La_{1-x}Sr_xNiO_2~~(x\simeq 0.2)$ which too showed insulating behaviour below 150 K.  In an attempt to increase electronic bandwidth authors replaced bigger ion $La$ with a smaller ion $Nd$. This lead to smaller cell volume and more electronic bandwidth. The system $NdNiO_2$ becomes the parent compound of $Nd_{0.8}Sr_{0.2}NiO_2$ which showed superconductivity from 9 K to 15 K.  However, more deeper understanding is required why $Nd_{0.8}Sr_{0.2}NiO_2$ is superconducting while $La_{0.8}Sr_{0.2}NiO_2$ is not (both thought to have $Ni ~~d^9$ configurations).

\section{A list of plausible explanations}
In the present note we would like to comment about the behaviour of resistivity of the parent compound and on the mechanism of superconductivity in the doped system. Resistivity of the parent compound $NdLiO_2$ is reproduced in figure (1).  It decreases with decreasing temperature upto $60~K$, and then shows upturn at lower temperatures. There are several possibilities for the observed resistivity minimum (figure 1) at around $60~K$ and up-turn at lower temperatures in $NdNiO_2$:

\begin{figure}[!h]
\begin{center}
\includegraphics[height=5cm,width=7cm]{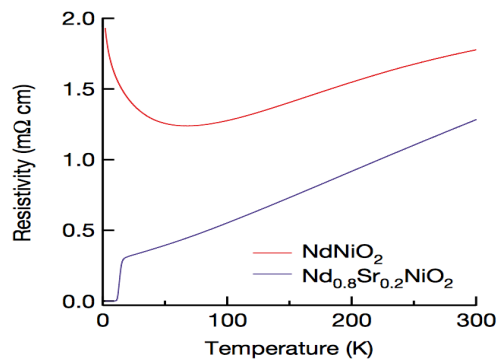}
\caption{Resistivity as a function of temperature. Upper red curve is for $NdNiO_2$. From Ref\cite{dang}.}
\end{center}
\end{figure}

%%%%%%%%%%%%%
\begin{itemize}

\item Weak localization effects: This is unlikely, as the system ($NdNiO_2$) prepared is reasonably good crystalline material (figure 1 and 2 in ref\cite{dang}).

\item As suggested in\cite{saw,lee}, 5d states of Neodymium atoms could be involved in electrical transport and there could be a coupling of these degrees of freedom with un-paired electrons in $Ni^{1+}$ leading to {\it Kondo system like behaviour} and resistivity minimum. This is quite plausible.  Experiments to prove or disprove this explanation are proposed at the end of this section.

\item The third possibility is temperature driven intra-band transitions. The resistivity upturn starts at around $60~K$ and at roughly  the same temperature Hall coefficient turns large negative in the undoped system $NdNiO_2$, and changes sign from being negative to positive in the doped system. The underlying cause could be analyzed in the following way:

At zero doping and zero temperature, $Nd~5d_{xy}$ bands form mobile states. There is direct $Nd~5d_{xy}-Nd~5d_{xy}$ overlaps and mobile bands are formed\footnote{LDA calculations show\cite{lee} that $Nd~5d$ bands are quite broad (from -0.2 eV to 8 eV)}, whereas $Nd5d_{3z^2-r^2}-Ni 3d_{3z^2-r^2}$ hybridization form localized states for the ab-plane electrical conduction (i.e., transverse direction to this hybridization state). There exists finite crystal field splitting  between the mobile states and localized states.

There are another important degrees of freedom central to superconductivity.  It is the hybrid band $Ni3d_{x^2-y^2}-O2p$ which exists in the ab-plane  and does not support any mobile state.  The reason is the standard one: double occupancy at $Ni 3d_{x^2-y^2}$ site leads to large Hubbard U.\footnote{It has been pointed out that Nickelates are Mott insulators while Cuprates are charge transfer insulators\cite{jiang}}

At finite temperatures the mobile $Nd~5d$ states are more populated and provide the dominant contribution to conduction in the ab-plane. As the temperature is reduced below $60~K$, the electronic population of the conducting $Nd~5d$ band reduces and that of the hybrid states $Nd5d_{3z^2-r^2}-Ni 3d_{3z^2-r^2}$ improves (we discuss other possible experiments to detect a signature of this splitting ($60~K \simeq 5~meV$)).  Thus, the resistivity up-trun in the undoped compound below $60~K$ points towards {\it the loss of carriers in the conducting band. This loss of carriers is also visible in the Hall coefficient data which drops towards large negative values as shown in figure (2) [red curve]}. As the Hall coefficient is given by:

\beq
R_H(T) = - \frac{1}{n(T) e c},
\eeq
the decrease in $n(T)$ with lowering temperature in mobile band $Nd~5d$ leads to larger negative Hall coefficient( figure (2) lower curve).

\begin{figure}[!h]
\begin{center}
\includegraphics[height=5cm,width=7cm]{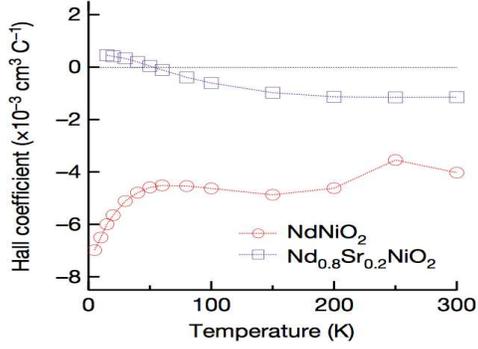}
\caption{Hall coefficient as a function of temperature. Lower red curve is for $NdNiO_2$. From Ref[\cite{dang}]}.
\end{center}
\end{figure}

Thus in the undoped system $NdNiO_2$ conductivity is through $Nd~5d$ bands, and contribution to conductivity of the hybrid band $Ni3d_{x^2-y^2}-O2p$ (which is the base of superconductivity) is zero due to Mott physics.

Next  question is what happens on doping. When the system $Nd NiO_2$ is doped with Strontium ($Nd_{0.8}Sr_{0.2}NiO_2$)  holes are created in $Ni 3d_{x^2-y^2}$ states (a small hole pocket is opened up\cite{yus,hep,hir,hiro,bot,xia}). The hole doping leads to the removal of the un-paried electron from $Ni3d_{x^2-y^2}$ states, leaving $Ni^{2+}$. Thus a typical charge configuration on Ni will be a mixture of $d^9$ and $d^8$ states. With Sr concentration 0.2 in the above system, 20 percent Ni atoms in a given sample will be in $d^8$ configuration and 80 percent will be having $d^9$ configurations. But the situation is not static.  Even at zero temperature, configuration at a given Ni atom will fluctuate between $d^8$ and $d^9$ as electrons flit from one Ni atom to another. This is due to quantum fluctuations. And the hybrid band $Ni 3d_{x^2-y^2}-O2p$ opens out another conducting channel  (hole pocket) {\it along with} the existing $Nd~5d$ one (electron pocket). This is like two resistances working in parallel (figure 3), and leads to even lesser resistivity.
\begin{figure}[!h]
\begin{center}
\includegraphics[height=5cm,width=6cm]{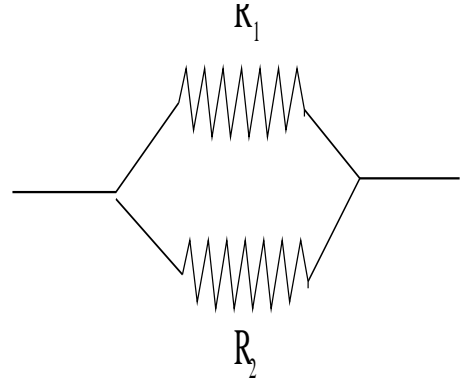}
\caption{Parallel resistances constituted by mobile $Nd5d$ states ($R_1$) and hole doped $Ni3d_{x^2-y^2}-O2p$ hybrid states ($R_2$), leading to net lower resistivity of the doped compound $Nd_{0.8}Sr_{0.2}NiO_2$.}
\end{center}
\end{figure}
Reduced resistivity can be seen in figure (1) for the doped system case. Conduction is electron like in the band $Nd5d$ whereas the conduction is hole like in the band $Ni3d_{x^2-y^2}-O2p$ . Thus at around $60~K$ the conduction happens via the doped holes in $Ni3d_{x^-y^2}-O2p $ band where it is hole-like and leads to sign change of the Hall coefficient from negative to position at that temperature (figure 2). This is one possible explanation of the observed behaviour.

We return back to the "Kondo system like behaviour" possibility. If 5d states of Neodymium atoms are involved in electrical transport and there is a coupling of these degrees of freedom with un-paired electrons in $Ni^{1+}$, then it can lead to {\it Kondo system like behaviour} and resistivity minimum can be understood in the undoped system\cite{saw,lee}. Now, when the system is doped, the doped holes goes to $Ni3d_{x^2-y^2}-O2p$ strongly correlated band. It could have two consequences: (1) It weakens the Kondo Coupling, and (2) another conducting channel is opened up (our parallel resistance idea (figure 3)). The result is the reduced resistivity of the doped system $Nd_{0.8}Sr_{0.2}NiO_2$ (figure 1) which on further lowering the temperature "skips" the Kondo up-turn and the system goes into the superconducting state. But how to reconcile the behaviour of the Hall coefficient is a challenging open issue in this scenario.

\item There is yet another possibility. It is known that Nickelates support various charge density wave orders ($NdNiO_3$ shows long range charge order below the metal-insulator transition at $T_{MI = 205~K}$\cite{garcia}). Thus there is a plausibility that $NdNiO_2$ also supports some form of charge order. There are two further possibilities:

Possibility No. 1: In the undoped system ($NdNiO_2$), if charge order happens in the $Nd~5d$ channel at around $60~K$, then due to the build up of the charge order, resistivity starts to increase as the temperature is reduced below $60~K$ (figure 1, upper curve). And when the system is doped, second conducting channel is opened up in $Ni3d_{x^2-y^2}-O2p$ strongly correlated band (as holes reside on $Ni$ in $Nd_{0.8}Sr_{0.2}NiO_2$). This "by-pass"  or "shunt" the higher resistive channel (figure 3) of CDW order and net resistivity is less for the doped system (figure 1, lower curve).  The behaviour of the Hall coefficient can also be reconciled in the following way. Louis Taillefer and collaborators has done extensive amount of experimental investigations in the case of charge order in under-doped Cuprates. Along with other probes they observe simultaneous rise in resistivity and fall in the Hall coefficient when temperature is lowered below $T_{Charge~Order}$\cite{taillefer}. These connections are quite intriguing! And much work is needed in this direction. However, it is difficult to reconcile the change in the Hall coefficient from negative to positive at $60~K$ (figure (2) upper curve) in this "picture".

Possibility No. 2: There is a possibility of self doping even in the case of $NdNiO_2$. Then along with configurations of the type $Nd^{3+}-Ni^{1+}$, there will also be configurations of the form $Nd^{2+}-Ni^{2+}$. This implies that even in the undoped system $NdNiO_2$, in the strongly correlated band $Ni3d_{x^2-y^2}-O2p$, there are holes on $Ni$ ions. These holes destroy Mott insulation, and the system is conducting via two channels: $Nd-Nd$ direct channel and $Ni3d_{x^2-y^2}-O2p$. Now, holes in the undoped system in the $Ni3d_{x^2-y^2}-O2p$ hybrid states are not randomly distributed, but they organize in the form of stripes, at and below the temperature $T_{Stripes} = 60~K$. This can lead to resistivity up-turn of the undoped system below $60~K$! And when the system is further hole doped with $Sr$, charge order temperature is shifted to even lower temperatures. But before that stipulated temperature the doped system undergoes superconducting transition.

These are some of the possibilities to comprehend the observed behaviour. The field of Nickelate superconductivity is at a very early stage of development. With more experiments in future the "actual picture" will emerge. Actually, all the experimental arsenal used to analyze Cuprates can be used in the case of Nickelates. Below we suggest some of the experiments that will be useful to resolve the cases:

\begin{enumerate}
\item To analyze CDW order, for example, Seebeck coefficient and Nernst coefficient can be measured across the $60~K$ mark.
\item Heat capacity measurements and quantum oscillations measurements can tell about the presence or absence of heavy fermions and Kondo physics in Nickelates.
\item Heat capacity measurements can detect the $5~meV$ gap (as the itinerant degrees of freedom start to freeze below $60~K$, and $\gamma~T$ term in the heat capacity will reduce).
\end{enumerate}

\end{itemize}
%%%%%%%%%%%%%%%

\section{Unconventional superconductivity}

The conducting hole doped band ($Ni3d-O2p$) in $Nd_{0.8}Sr_{0.2}NiO_2$ could be responsible for superconductivity. There are three possibilities to rationalize superconductivity in these systems:

\begin{enumerate}
\item Unconventional superconductivity from repulsion: This is the most favorable case. A sign changing gap like d-wave is expected\cite{xia,kei}.

\item Superconductivity from magnetic spin fluctuations as glue. Magnetic spin fluctuation glue is unlikely way to understand Nickelate superconductivity.  There are at least two reasons: (1) these do not showing any magnetic instability and (2) it is difficult to imagine same electrons providing glue and the same electrons participating in pairing as argued in ref\cite{kei} (as in the case of Cuprates). One possible  reason why $NdNiO_2$ does not show AFM order could be that it has much weaker superexchange interaction $J$\cite{jiang}.

\item Plain vanilla RVB approach\cite{ander}: This is an alternative possibility. Nickelates can be thought of as Cuprates but with reduced correlation effects. Mott physics is not observed in the case of Nickelates. The reason is due to the possibility of "by-passing" or "short circuiting" the quantum jam in $Ni3d_{x^2-y^2}-O2p$ states via $Nd$ conducting channel. But this does not stop one to use $t-J$ model for Nickelates also ($Ni3d_{x^2-y^2}-O2p$ states).  Basically, at zero doping and zero temperature one has un-paired localized electrons in the  $Ni3d_{x^2-y^2}-O2p$ hybrid band. Thus Hubbard model based approaches become applicable to this  case also. Therefore, Mott physics based approaches may be applicable to Nickelates.  But, there are  problems with the Plain vanilla RVB approach even in the case of Cuprates\cite{verma}: (1) Pairing is assumed a-priori through superexchange interaction J and (2) RVB approach gives continuous behaviour across the quantum critical point $p^*$ where pseudogap boundary ends\cite{verma}. Experimentally, it found that at $p^*$ several observables show anomalous behaviour (like heat capacity $\propto -lnT$, Hall coefficient etc.)\cite{tail}. If it turn out that Nickelates do not show quantum criticality then RVB-based approach is even more appropriate. On the other hand if quantum criticality is found in Nickelates, then the mainstream approach "superconductivity from repulsion" will be more appropriate. 

\end{enumerate}

Thus, superconductivity in Nickelates can be rationalized through point (1). In this case also (repulsive mechanism for unconventional superconductivity) {\it we do not have a predictive theory} as is well known in the case of Cuprates.

We end this note with some more comments:

\begin{itemize}
\item More experimental work related to study of superconductivity at various  doping levels is much needed (which is extremely difficult in Nickelates). Question is whether there is a dome type structure?

\item Whether there is a quantum critical point at a special doping level of $Sr$? Measurement of heat capacity and hall coefficient at that point can establish more connections with cuprates (or can rule out).

\item Heat capacity/Quantum oscillation measurements to rule out (or in) the heavy fermion behaviour (as suggested in ref\cite{saw}). 

\end{itemize} 

Actually, all the questions which were asked in Cuprate superconductivity can be asked in the case of Nickelates. Clear differences are (1) the lack of magnetism, and (2) the role of spacer layer (Nd layer) in electrical conduction, in the case of prepared Nickelate compounds.

%----------------------------------------------------------------------------------------
%	REFERENCE LIST
%----------------------------------------------------------------------------------------

%----------------------------------------------------------------------------------------

\end{document}